\documentclass[a4,aps,amsmath,floatfix,nofootinbib,10pt]{revtex4} 
\usepackage{graphicx} \usepackage{color} \usepackage{enumerate} 
\newcommand{\be}{\begin{equation}} \newcommand{\ee}{\end{equation}} 
\newcommand{\ba}{\begin{array}} \newcommand{\ea}{\end{array}} 
\newcommand{\bea}{\begin{eqnarray}} \newcommand{\eea}{\end{eqnarray}} 
\newcommand{\bdm}{\begin{displaymath}} 
\newcommand{\edm}{\end{displaymath}} 
 
\newcommand{\sgn}{\operatorname{sgn}} 
\newcommand{\erf}{\operatorname{erf}} \begin{document}

\title{Hysteresis in the zero-temperature random-field Ising model on 
directed random graphs.}

\author{Prabodh Shukla}

\affiliation{North Eastern Hill University \\ Shillong-793 022, India}

\begin{abstract}

We use zero-temperature Glauber dynamics to study hysteresis in the 
random-field Ising model on directed random graphs. The critical 
behavior of the model depends on the connectivity $z$ of the graph 
rather differently from that on undirected graphs. Directed graphs and 
zero-temperature dynamics are relevant to a wide class of social 
phenomena including opinion dynamics. We discuss the efficacy of 
increasing external influence in inducing a first-order phase 
transition in opinion dynamics. The numerical results are supported by 
an analytic solution of the model.

\end{abstract}

\maketitle

\section{Introduction}

Extensive quenched disorder in a thermodynamic system endows its free 
energy landscape with an abundance of local minima separated by high 
energy barriers~\cite{young}. This prevents the system from relaxing to 
its ground state over practical time scales. Therefore the focus of 
experimental observations and theoretical models shifts to 
nonequilibrium effects~\cite{bertotti}. The random-field Ising model 
(RFIM)~\cite{imryma} is perhaps the simplest model of a system with 
quenched disorder. It models the disorder by on-site random fields. 
Some other popular models e.g. the Sherrington-Kirkpatrick 
(SK)~\cite{sherrington} model of spinglass, or the Hopfield 
model~\cite{hopfield} of a neural network are based on random bonds. 
{\color{blue} {An analytic solution of models with quenched site or 
bond disorder has proven elusive. Our understanding of their behavior 
is mainly based on numerical simulations. A few analytic results are 
available in the case of site disorder. For example, it was proven 
after considerable debate that the lower critical dimension for RFIM is 
equal to two ~\cite{imbrie}. As far as we know, a result with similar 
rigor is not available in the case of bond disorder.}} In general, the 
simplicity of Ising models is deceptive. As is well known, the parent 
Ising model without quenched disorder has remained unsolved so far 
except in one dimension ~\cite{ising}, and on a square lattice in the 
absence of an applied field ~\cite{onsager}. {\color{blue} {A few exact 
results are also available for honeycomb, triangular, and certain other 
nets ~\cite{wannier, syozi} but the three dimensional case remains a 
distant dream.}} We have to resort to numerical simulations for 
questions of general interest. Simulations too suffer from a similar 
difficulty as do the experiments. The system fails to equilibrate on 
realistic time scales. Initially the RFIM as well as the SK model were 
proposed to understand the equilibrium behavior of quenched disordered 
systems. After more than a decade of intense activity in this 
direction, the effort vaned somewhat. {\color{blue} The reason could be 
that the basic physics of the model came to be understood reasonably 
well but an exact solution on periodic lattices seemed too difficult}. 
However it did contribute a great deal to our understanding of 
disordered physical and social systems and gave birth to new ideas. 
Efforts towards a correct solution of the SK model in the mean field 
limit lead to the idea the replica symmetry breaking 
~\cite{parisi,mezard}. The RFIM returned to a second innings of play 
when it was used along with Glauber dynamics ~\cite{glauber} to study 
hysteresis and Barkhausen noise in disordered ferromagnets at zero 
temperature and in the limit of zero frequency of the driving field 
(ZTRFIM)~\cite{sethna1}. Although these limits are unattainable in a 
real experiment, the model can be solved exactly on a Bethe lattice and 
produces interesting nonequilibrium critical behavior ~\cite{dhar}. 
This provides a reasonable understanding of the observed phenomena in 
hysteresis experiments ~\cite{sethna2}.

We recapitulate briefly the key results and issues related to the 
solution of ZTRFIM on a Bethe lattice of connectivity $z$ and nearest 
neighbor ferromagnetic interaction $J$. The solution exhibits hysteresis 
for all $z\ge2$. Moreover each half of the major hysteresis loop has a 
discontinuity at applied fields $h=-h_c$ and $h=h_c$ respectively if $z
> 3$ and the standard deviation $\sigma$ of the Gaussian random field
is less than a critical value $\sigma_c$. The critical value depends on 
$z$; $\sigma_c\approx 1.78J$ for $z=4$ and increases with increasing 
$z$. The discontinuity in magnetization occurs at $|h_c| > J$ and with 
increasing $\sigma$ it moves towards $|h_c| \to J$ where it vanishes 
continuously as $\sigma \to \sigma_c$ from below. {\color{blue}{The 
point $\{h_c=J,\sigma=\sigma_c\}$ is a nonequilibrium critical point.}} 
Various thermodynamic quantities in the vicinity of this point show 
universality and scaling as observed near the critical temperature of 
pure Ising model. These results pose two puzzles: (i){\color{blue} {why 
disorder-driven hysteresis in the limit of zero temperature and zero 
frequency of the driving field should show a critical point with 
scaling and universality properties quite similar to those at the 
temperature-driven equilibrium critical point,}} and (ii) why should 
nonequilibrium critical behavior occur only if $z > 3$ while in 
equilibrium it occurs for $z>2$. There have been several attempts to 
understand the physics behind this including a mapping of the problem 
to a branching process in population dynamics ~\cite{handford}. Further 
clarity emerged when the analysis for integer values of $z$ was 
extended to continuous vales of $z$. Of course the connectivity of a 
node is necessarily integer but it could be distributed over a set of 
integers to make the average connectivity a continuous variable. The 
continuous $z$ model has been solved in two cases ~\cite{shukla}, (i) 
mixed lattice with each node having $z=4$ with probability $c$ and 
$z=3$ with probability $1-c$ and (ii) randomly diluted $z=4$ lattice 
with a fraction $c$ of the nodes occupied and $1-c$ unoccupied. Note 
that the connectivity of occupied nodes on the diluted lattice could be 
$z=1,2,3,4$. On the mixed lattice, it was shown that critical 
hysteresis occurs for all $c>0$. On the randomly diluted lattice, it 
occurs only if $c \ge \frac{2^{1/3}}{1+2^{1/3}} \approx 0.5575$. Taking 
both cases into account, it was resolved that critical hysteresis 
requires two conditions, (i) a spanning path across the lattice and 
(ii) an arbitrarily small sprinkling of $z\ge4$ nodes on this path. 
Nodes with $z\ge4$ are essential for an infinite avalanche. They serve 
to boost the avalanche along its path and ensure it does not die in its 
tracks. A diverging correlation length in equilibrium does not have 
similar issues. An arbitrary small fraction of $z=4$ sites on a 
spanning path has important bearings on the complexity of the model as 
well~\cite{rosinberg}. We have a through understanding of various 
technical issues related with the ZTRFIM on a Bethe lattice. However 
its application to magnetic hysteresis has a disconcerting aspect. The 
key feature of ZTRFIM is the occurrence of a critical point in the 
model. The critical point disappears when conditions (i) temperature $T 
\to 0$ and (ii) time period of observation $t \to \infty$ are relaxed. 
These conditions have to be relaxed in experiments on magnets.

The ZTRFIM may be applied to study hysteresis in social systems as 
well, e.g. opinion dynamics~\cite{sirbu,quattrociocchi}. Suppose each 
Ising spin represents an individual (agent) in a society where only two 
brands of tooth pastes are sold, $A$ or $B$. Initially everybody uses 
$A$ because it is the only brand available. Then $B$ is introduced 
accompanied by a mass media advertisement to switch to it. 
Advertisement is like the field $h$ because all agents are exposed to 
it uniformly. We take the range of $h$ to be $[-\infty,\infty]$ as in 
ZTRFIM. In the absence of interactions between agents, we may expect 
the fraction of agents using $B$ to increase gradually from zero to 
unity as $h$ is ramped up from $-\infty$ to $\infty$. No hysteresis is 
expected in this case i.e. the fraction of agents using $A$ or $B$ at 
$h=0$ is expected to be equal to half. Now we expose each agent $i$ to 
a quenched random field $h_i$ and a field $(2z-n)J$ from its $z$ 
"friends" with whom it communicates; if $h_i<0$ the agent has an innate 
preference for brand $A$, otherwise for band $B$; $(2z-n)J$ is the 
influence on the agent from its $z$ friends, $n$ of which use brand $B$ 
at a given $h$. As we know, there is hysteresis in this case if $z\ge2$ 
and if friends are friends of each other (undirected graph). The 
fraction of agents using $B$ is influenced by the memory of what brand 
they used earlier. If $z\ge4$, and the variation in individual 
preferences $\sigma$ is small, there is a sharp jump in the sale of 
brand $B$ at a particular $h$. It is clearly a very useful information 
for the advertiser who wishes to maximize her profit against the cost 
of advertising.

In this paper we modify the interactions in the ZTRFIM in the context 
of opinion dynamics. We allow the possibility that "friends" need not 
be friends of each other. Equivalently, edges on the graph need not be 
two-way streets. In other words, we consider directed graphs. In 
physical systems the interaction $J_{ij}$ between nearest neighbors $i$ 
and $j$ is necessarily symmetric i.e. $J_{ji}=J_{ij}$ but in social 
systems it is often not the case as in a boss and subordinate 
relationship. In contrast to the case of undirected graphs, directed 
graphs show no hysteresis if $z=2$ but hysteresis including critical 
hysteresis if $z \ge 3$. We examine partially directed $z=3$ graphs as 
well where a fraction $c$ of the nodes have two undirected and one 
directed edge, and the remaining fraction $1-c$ have all undirected 
edges. We derive analytical results for criticality on such graphs and 
verify these by simulations as discussed below.

\section{The Model, Simulations, and Theory}

Consider Ising spins $\{s_i=\pm1\}$ situated on $N$ nodes $\{i=1,2, 
\ldots,N\}$ of a random graph. Each node is randomly linked to $z$ 
other nodes. Each link \{i,j\} may be a one way or a two way street. If 
the node $j$ can influence node $i$ but node $i$ can not influence node 
$j$, we call the edge $J_{ij}$ directed and set $J_{ij}=J$ and 
$J_{ji}=0$, otherwise $J_{ij}=J_{ji} =J >0$. We consider graphs which 
have a mixture of directed and undirected edges. The Hamiltonian of the 
system is,

\bdm H=-\sum_{i,j}J_{ij}s_i s_j-\sum_i h_is_i-h \edm

$J$ sets the energy scale, $h_i$ is a Gaussian quenched random field 
with average zero and variance $\sigma^2$, $h$ is a uniform applied 
field measured in units of $J$. The system evolves under zero 
temperature Glauber dynamics. A node $i$ is selected at random and the 
net local field on it $\ell_i=-\sum_j J_{ji}s_j-h_i -h$ is evaluated. 
The spin $s_i$ is flipped if $s_i\ne \sgn{\ell_i}$. The procedure is 
repeated at a fixed $h$ until all spins get aligned along the net 
fields at their site. The hysteresis loop is obtained as follows. We 
start with a sufficiently large and negative $h$ such that the stable 
configuration has all spins down $\{s_i=-1\}$. Then $h$ is increased 
slowly till some spin flips up. It generally causes some neighboring 
spins to flip up in an avalanche. We keep $h$ fixed during the 
avalanche and calculate the magnetization $m(h)=\sum_is_i/N$ after the 
avalanche has stopped. The procedure is repeated till all spins in the 
system are up; $m(h)$ at increasing values of $h$ between avalanches 
makes the lower half of the hysteresis loop. The upper half is obtained 
similarly. As is well known, the model shows hysteresis for $z\ge2$ on 
undirected graphs. The situation is different on a directed graph where 
each edge between a pair of nodes can pass a message in one direction 
only.

Fig.1 shows hysteresis loops on directed graphs of connectivity 
$z=2,3$. We focus on $z\le3$ because we find the case $z > 3$ to be 
qualitatively similar to $z=3$. The figure shows absence of hysteresis 
for $z=2$ and $\sigma=1$ i.e. $m(h,\sigma=1,z=2)$ in increasing $h$ 
coincides with the one in decreasing $h$. Although only one value of 
$\sigma$ is shown in Fig.1 for $z=2$ but qualitatively similar result 
is obtained for all $\sigma >0$. The is in contrast to the behavior on 
an undirected Ising chain which shows hysteresis but no critical points 
on the hysteresis loop. The physical reason for the absence of 
hysteresis on the $z=2$ directed chain is not immediately obvious but 
becomes clear when equations for the loop are considered. We write the 
equations for general $z$ and discuss $z=2$ and $3$ as special cases. 
The lower and upper halves of the loop are related to each other by 
symmetry. Therefore it suffices to focus on the lower half. Initially 
the probability $P(h=-\infty,\sigma;z)$ that a randomly chosen node of 
connectivity $z$ is up is zero. When the system is exposed to a field 
$h$ and relaxed, spins flip up in an avalanche and $P(h;\sigma;z)$ 
increases with each iteration $t$ of the dynamics until it reaches a 
fixed point value $P^*(h;\sigma;z)$. The evolution is governed by the 
equation,

\be P^{t+1}(h;\sigma;z)=\sum_{n=0}^z {z\choose n} [P^t(h;\sigma;z)]^n 
[1-P^t(h;\sigma;z)]^{z-n}p_n(h;\sigma;z), \ee 

where $p_n(h;\sigma;z)$ is the probability that the random field $h_i$ at 
a node is large enough such that it is up if $n$ of the $z$ neighbors 
which are linked to it are up at an applied field $h$.

\be p_n(h;\sigma;z)=\frac{1}{\sqrt{2 \pi \sigma^2}} 
\int_{(z-2n)J-h}^{\infty} e^{-\frac{h_i^2}{2 \sigma^2}}dh_i \ee

The rationale behind equation (1) is the $z$ neighbors which affect the 
state of their common node are themselves not affected by it. The 
magnetization in a stable state is given by the equation 
$m^*(h;\sigma;z) = 2 P^*(h;\sigma;z)-1$. It is easily verified that 
equation (1) has the symmetry $m(h;\sigma;z)=-m(-h;\sigma;z)$ and 
therefore $m(h=0;\sigma;z)=0$ or equivalently $P^*(h=0;\sigma;z)=1/2$ 
is always a solution of equation (1) for any $\sigma$ and $z$. However 
it can become unstable depending on $\sigma$ and $z$. The stability 
analysis of $m^*(h=0;\sigma;z)=0$ in the linear approximation reveals 
that a perturbation $\delta m_0$ to it transforms to $\delta m_1 = 
A(\sigma;z) \delta m_0$ under the next step of the dynamics. We obtain 
for $z=2,3$ respectively,

\be A(\sigma;z=2)=\left[ \erf{\frac{2J}{\sqrt{2 \sigma^2}}}\right] \ee

\be A(\sigma;z=3)=\frac{3}{4} \left[ \erf{\frac{3J}{\sqrt{2 \sigma^2}}} 
+ \erf{\frac{J}{\sqrt{2 \sigma^2}}}\right] \ee

For finite $\sigma$ and $J$, $A(\sigma;z=2) < 1$. Thus perturbations 
decrease to zero under repeated applications of the iterative dynamics. 
In other words, $m^*(h=0;\sigma;z=2)=0$ is stable and there should be 
no hysteresis on the $z=2$ lattice for any finite $\sigma$ as indeed 
seen in Fig.1. For $z=3$, the fixed point $m^*(h=0;\sigma;z=3)=0$ is 
stable only if $\sigma > \sigma_c$ where $\sigma_c$ is determined by 
the equation $A(\sigma_c;z=3)=1$. This gives $\sigma_c \approx 1.781 
J$. For $\sigma < \sigma_c$ the unstable fixed point 
$m^*(h=0;\sigma;z=3)=0$ bifurcates into two stable fixed points, one 
negative and the other positive, resulting in magnetization reversal 
with a jump at some $\sigma$-dependent applied field $h$ on the 
hysteresis curve. Fig.1 compares the exact solution presented above 
with simulations for three representative values of $\sigma=J, 1.5 J, 2 
J$ for $z=3$. {\color{blue}{ As may be expected, the fit between theory 
and simulations is excellent. The simulations were performed on a 
system of size $N=10^6$ for a single configuration of the random-field 
distribution. The results are indistinguishable from the corresponding 
theoretical results on the scale of the figure. The agreement between 
theory and simulations remains good in closer vicinity of $\sigma_c 
\approx 1.781$ as well but it is not shown in Fig.1 in order to avoid 
crowding the figure.}}

It is also of interest to consider partially directed random graphs. We 
consider a partially directed graph with connectivity $z=3$. A fraction 
$c$ of the nodes have two directed and one undirected edge. All three 
edges of the remaining fraction $1-c$ are directed. Analytical results 
for this case are presented in the following. Theoretical and 
simulation results are compared in Fig.2 for a few representative 
values of $c$ and $\sigma=1.5 J$. For $c=0$, we of course recover the 
results depicted in Fig.1. As $c \to 1$, the hysteresis loops widen and 
the first-order jumps in $m(h;\sigma;z)$ decrease in size with 
increasing $c$. The jumps may persist at $c=1$ if $\sigma < \sigma_c$. 
We find $\sigma_c \approx 1.522 J$ and $h_c \approx 0.281 J$ at $c=1$. 
As $c$ increases from $c=0 \to 1$, $\sigma_c$ decreases from $1.781 J 
\to 1.522 J$ and the critical field at which the jump vanishes shifts 
from $h_c=0 \to h_c\approx 0.281 J$. To obtain the theoretical 
expression for the hysteresis loop we focus on the lower half of the 
loop. Let $P^{t}$ be the probability that a randomly selected site is 
up at the $t$-th iteration of the dynamics, and $P^t_c$ be the 
conditional probability that a neighbor of a yet unrelaxed site is up. 
We have suppressed the arguments of the probability functions to 
simplify the notation. Starting from the initial state $P^0_c=0$ and 
$P^0=0$, the update rules for the coupled probabilities are,

\be P^{t+1}_c=(1-c^2) P^t + c^2 \left[ p_2 \{P^t\}^2 +2 p_1 \{P^t\} 
\{1-P^t\} + p_0 \{1-P^t\}^2 \right] \ee

\be P^{t+1}=\{P^t\}^2 [p_3 P^t_c +p_2 (1-P^t_c)]
                   +2 \{P^t\} \{1-P^t\} [p_2 P^t_c + p_1 (1-P^t_c)]
                   +\{1-P^t\}^2 [p_1 P^t_c + p_0 (1-P^t_c) ] \ee

The above equations are understood as follows. At step $t+1$ we need to 
consider only the sites that are down because those which have already 
turned up do not turn down again. Choose a down site $D$ and let its 
neighbors $A_1$, $A_2$, $A_3$ be linked to $D$ by edges $DA_1$, $DA_2$, 
$DA_3$. Choose a neighbor at random, say $A_1$. The probability that 
$A_1$ is up before $D$ is relaxed depends on whether the edge $DA_1$ is 
directed or not. The edge $DA_1$ is directed with probability $1-c^2$. 
In this case $D$ has no influence on $A_1$ and the probability that 
$A_1$ is up is equal to $P^t$. This accounts for the first term in 
equation (5). The second term in equation (5) pertains to the case when 
$DA_1$ is undirected. Note that if $DA_1$ is undirected, the other two 
edges meeting at $A_1$ must be directed. With similar reasoning, 
equation (6) gives the probability that $D$ flips up when relaxed. 
Terms in square brackets refer to configurations of $A_1$, and those in 
curly brackets to configurations of $A_2$ and $A_3$. Equations (5) and 
(6) are iterated till a fixed point is reached. Magnetization in the 
fixed point state is given by $m^*(h,\sigma,c)=2 P^*(h,\sigma,c)-1$. 
Fig. 2 shows the theoretical result for $\sigma=1.5 J$ and $c=0.00, 
0.75, 1.00$ along with the corresponding simulation results for 
comparison. As may be expected, the fit is excellent.

\section{Discussion}

We have presented an analytic solution of ZTRFIM on directed graphs and 
verified the solution in special cases by numerical simulations. The 
availability of an analytic solution is clearly valuable for 
understanding phase transitions in a system. Although directed graphs 
may not be of direct relevance to physical systems but they have been 
used to study social phenomena including opinion dynamics. To the best 
of our knowledge, bulk of the work on opinion dynamics has been carried 
out in the absence of an external influence. Hysteretic effects have 
received relatively little attention. We are not aware of appropriate 
field data that can be used to test the predictions of the model 
presented here. However qualitative predictions appear to bear out our 
experience with the remarkable effectiveness of advertisements. In a 
population where each person receives a nonreciprocal recommendation 
for a new product from three or more individuals, the sale of the 
product is predicted to shoot up sharply with a modest amount of 
advertisement and stay at a high level even after the advertisement is 
discontinued. The narrower is the variation in the individual 
preferences in the population, the stronger is the effectiveness of 
advertisement. These trends seem to be qualitatively true and may be 
further exploited in marketing a product.

\begin{figure}[ht] 
\includegraphics[width=0.75\textwidth,angle=0]{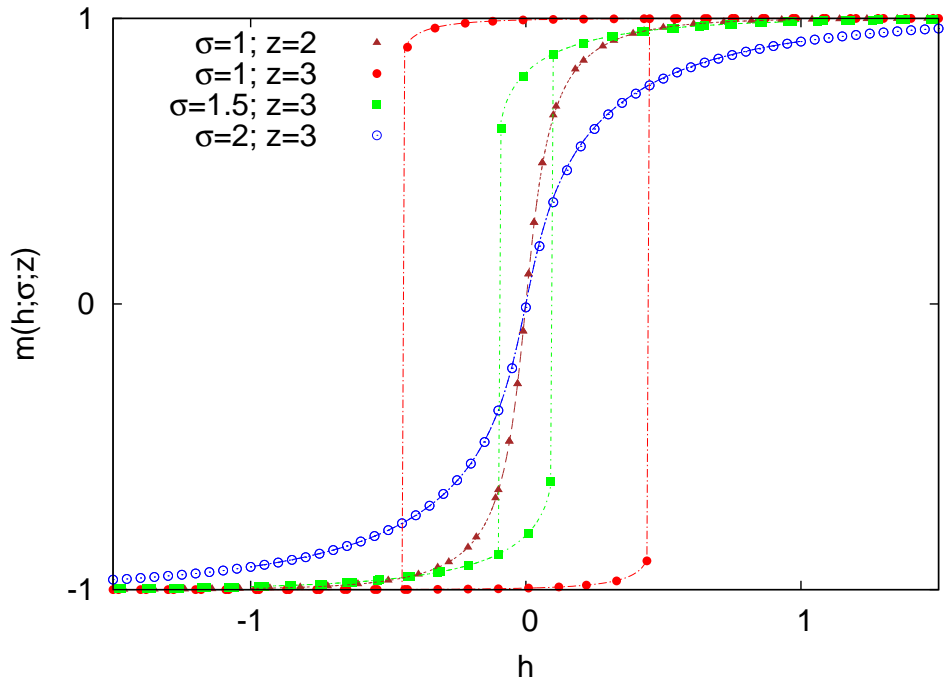}

\caption{ Hysteresis on directed random graphs of connectivity $z=2,3$ 
for a Gaussian distribution N(0,$\sigma^2$) of the random field. We 
have set $J=1$. Theoretical results are shown by continuous curves. 
Symbols depict results from numerical simulations of the model 
{\color{blue}{for a single configuration of the random-field 
distribution on a system of size $N=10^6$. Numerical results are 
indistinguishable from the theoretical results on the scale of the 
figure}}. Hysteresis is absent on a $z=2$ graph for any $\sigma$ and 
also on a $z=3$ graph if $\sigma > \sigma_c \approx 1.781$. For $\sigma 
< \sigma_c$ the loop has discontinuities at $h=\pm h_c$. The 
discontinuities reduce in size and move towards $h=0$ with increasing 
$\sigma$. As $\sigma \to \sigma_c$, the discontinuities vanish 
continuously at $h_c=0$.} \label{fig1} \end{figure}

\begin{figure}[ht] 

\includegraphics[width=0.75\textwidth,angle=0]{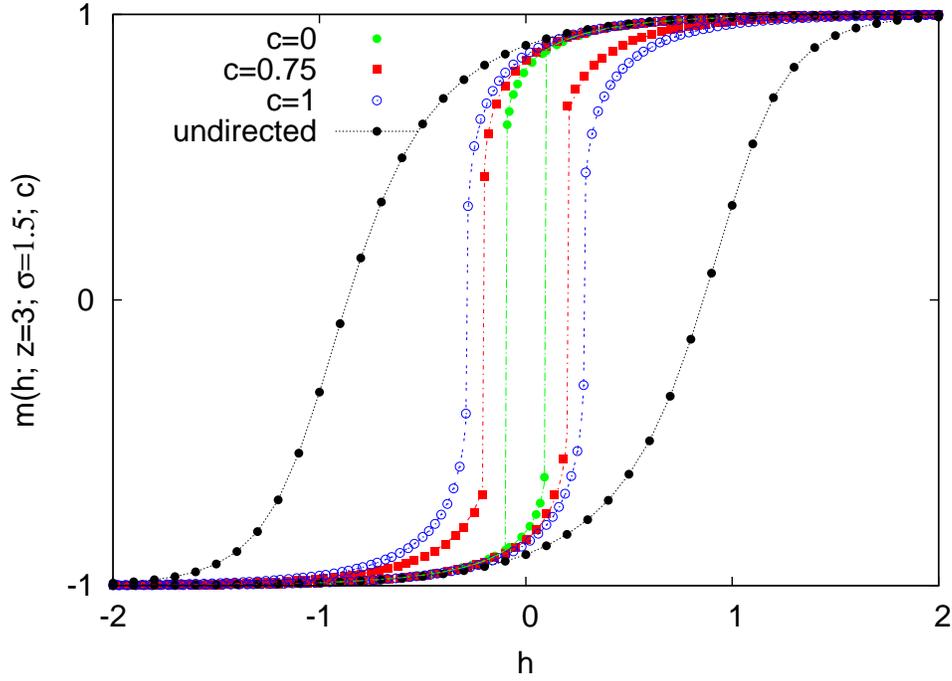} \caption{ 
Hysteresis on a partially directed random graph with $z=3$ and 
$\sigma=1.5$ in units of $J$. A fraction $c$ of the nodes have two 
directed edges and one undirected. All three edges of the remaining 
fraction $1-c$ are undirected. Theoretical results are shown by 
continuous lines and simulations {\color{blue}{on a $N=10^6$ graph}} by 
symbols. Vertical portions of curves denote discontinuities. Results 
shown in Fig.1 are recovered for $c=0$. As $c \to 1$, the first-order 
jump in $m(h)$ reduces and moves away from the origin. A case where all 
three edges are undirected is also shown for comparison.} \label{fig2} 
\end{figure}

\end{document}